\documentclass[twocolumn,showpacs,preprintnumbers,amsmath,amssymb]{revtex4}
\usepackage{amsmath}
\usepackage{graphicx}

\newcommand{\be}{\begin{equation}}
\newcommand{\ee}{\end{equation}}
\newcommand{\ket}[1]{\left| #1 \right\rangle}

\begin{document}

\title{Suspension of atoms and gravimetry using a pulsed standing wave}

\author{K. J. Hughes, J. H. T. Burke and C.A. Sackett}
\affiliation{Physics Department, University of Virginia, Charlottesville, VA 
22904}
\email{sackett@virginia.edu}
\date{\today}

\begin{abstract}
Atoms from an otherwise unconfined $^{87}$Rb condensate are shown to be 
suspended against gravity using repeated reflections from a pulsed optical 
standing wave.  Reflection efficiency was optimized using a triple-pulse
sequence that, theoretically, provides accuracies better than
99.9\%.  Experimentally, up to 100 reflections are observed, 
leading to dynamical 
suspension for over 100~ms.  The velocity sensitivity of
the reflections can be used to determine the local gravitational
acceleration.  Further, a gravitationally sensitive atom interferometer
was implemented using the suspended atoms, with packet coherence maintained
for a similar time.  These techniques could be useful
for the precise measurement of gravity when it is impractical to 
allow atoms to fall freely over a large distance.
\end{abstract}


\maketitle

Many of the recent advances in the physics of ultra-cold atoms have
relied on the use of magnetic and/or optical fields to confine
atoms and suspend them against gravity.  Typically, confinement and 
suspension go together \cite{Sackett06}, 
so that if one wishes to study unconfined
atoms, they must be in free fall.  Achieving long observation 
times therefore requires either a large drop distance
or a microgravity environment \cite{Vogel06,APB06,Dimopoulos07}.
Freely falling atoms are also used when measuring gravity in atom
interferometry experiments \cite{Berman97}, 
since any applied trapping forces would spoil the measurement accuracy.
The large drop distances that this requires are one limit to the
performance and applicability of the technique.

In this Letter, we demonstrate a new method for suspending ultra-cold
atoms that imposes negligible confinement and can be sufficiently precise to 
be used in a gravimetry experiment.  The suspension consists of
``bouncing'' the atoms repeatedly from a pulsed optical standing wave.
A similar 
idea was recently proposed by Impens {\em et al} \cite{Impens06}.  
In addition to 
implementing the suspension method, we also demonstrate 
a gravity-sensitive interferometer using the confined atoms.
Although we do not achieve significant precision here, 
improvements may permit gravimetry with precision comparable to that
achieved using falling atoms, but requiring a negligible drop distance.

Atoms have previously been suspended using multiple bounces from
evanescent waves \cite{Aminoff93},
magnetized surfaces \cite{Roach95},
light sheets \cite{Bongs99},
and magnetic fields \cite{Arnold02b}.
However, these techniques
introduce complex forces that significantly modify
the atomic wave function from what it would be in a freely falling frame, 
and that typically are not characterized well enough to 
permit accurate gravimetry.
Furthermore, these methods have been limited to only a few bounces,
while we have observed up to 100 bounces 
and expect more to be possible.  
Our method is related to the phenomenon of 
Bloch oscillations of atoms held in a static standing wave potential
\cite{Raizen97},
which has also been used to measure gravity \cite{Ferrari06}
but does not model freely falling atoms.

\begin{figure}
\begin{center}
  \includegraphics[width=3in]{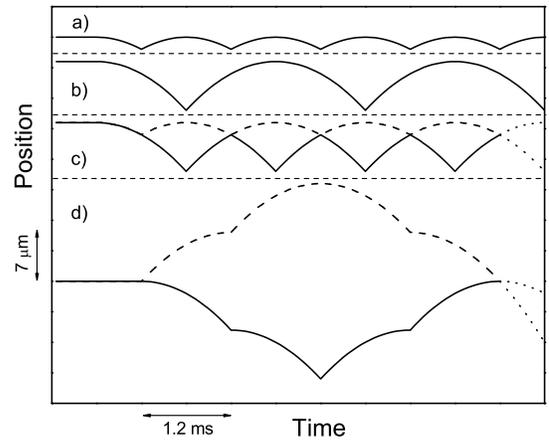}
  \caption{Atom trajectories discussed in this manuscript, shown with a 
  common scale as indicated.  
  (a) Bouncing atoms using order-1 reflections. (b) Bouncing atoms
  using order-2 reflections. 
  (c) Interferometer produced with a combination of order-1 and order-2
  reflections. Here the solid and dashed curves show the two arms of
  the interferometer, and the dotted curves show the possible outputs.
  (d) More complex interferometer with large arm spacing.
  Trajectories (a)--(c) were implemented here.}
  \label{fig:trajectories}
\end{center}
\end{figure}

The manipulation of atoms by an off-resonant standing-wave laser beam
was first used in thermal atomic beams, 
\cite{Martin88,Giltner95} and has 
proven useful for cold atoms as well \cite{Kozuma99,Wu05}.
Through the ac Stark effect, the laser induces a periodic potential 
that acts as a diffraction grating for the atomic wave function,
producing coupling between momentum components that differ by $2\hbar k$
for light with wavenumber $k$.
By controlling the intensity and duration of the applied pulse, various
beam-splitting and reflecting operations can be achieved.
Generally the results are sensitive to the initial velocity 
of the atoms, but the low velocity spread in an ultra-cold sample allows the
operations to be quite precise \cite{Hughes07}.

We define an order-$n$ reflection to be the operation driving 
$\ket{n\hbar k} \leftrightarrow \ket{-n\hbar k}$, where $\ket{p}$ denotes
an atomic state with momentum $p$.  Suspension of atoms using such an
operation starts with a sample of mass $m$ atoms held in a conventional trap. 
At time $t = 0$, the trap is switched off, allowing
the atoms to fall in the local gravitational field $g$.  At 
$t = t_n \equiv n\hbar k/mg$, the atomic momentum will 
reach $-n \hbar k$ and a
reflection operation is applied using a vertically oriented 
laser beam, reflecting the atoms upward with $p = +n\hbar k$.
They move ballistically for an interval $2t_n$, after which they again
have $p = -n\hbar k$ and the reflection can be repeated.  
We used both order-1 and order-2 reflections, with trajectories illustrated in
Fig.~1(a) and (b) respectively.

Losses during the reflection operation will limit the number of
bounces that can be achieved.  We studied a simple order-2 reflection
pulse in Ref.~\cite{Hughes07} and found it to be limited to a fidelity of 
0.94, which would permit roughly 20 bounces.  However, by shaping the
intensity-time profile of the light, we were able to achieve better
performance.  We modeled the operation by numerical solution of the
Schrodinger equation using the Bloch expansion
\be
\psi(z,t) = \sum_n c_n(t) e^{i(2nk + \delta)z}
\ee
for optical potential $V_L(z, t) = \hbar\beta(t) \cos(2kz)$.  Here $\psi$ is
the atomic wavefunction, 
$\delta$ accounts for an initial momentum offset, and $\beta(t)$ is
proportional to the light intensity.  This yields a set of equations
\be \label{eq-bloch}
i\frac{d c_n}{dt} = \frac{\hbar}{2m}(2nk+\delta)^2c_n + \frac{\beta}{2}
\left(c_{n-1}+c_{n+1}\right),
\ee
which were truncated at $n = \pm 6$.  For $\beta(t)$, 
we considered a symmetric sequence of
three square pulses with durations $(T_1, T_2, T_1)$ and intensities
$(\beta_1, \beta_2, \beta_1)$.  An optimization algorithm was used to determine
the best values of the $T$'s and $\beta$'s.  For the order-1 reflection,
we found $(T_1, T_2) = (0.355, 0.592) \omega_r^{-1}$, and
$(\beta_1, \beta_2) = (1.73, 3.45) \omega_r$, where
$\omega_r = \hbar k^2/(2m) \approx 2.36\times 10^4$~s$^{-1}$ is the 
atomic recoil frequency.  The order-2 reflection was optimized at
$(T_1, T_2) = (0.256, 1.46) \omega_r^{-1}$ and $(\beta_1, \beta_2)
= (2.28, 4.59) \omega_r$.  In both cases, the calculated fidelity exceeded
0.9998, but the model neglected losses due to spontaneous emission.
Maintaining an error below $5\times10^{-3}$ required 
$\delta/k$ to be less than $0.05$ for the order-1 operation and
$0.02$ for the order-2 operation. 

Bouncing was implemented using 
approximately $10^4$ $^{87}$Rb atoms from a 
Bose-Einstein condensate. The atoms were prepared in the
$\ket{F=2,m_F=2}$ hyperfine state in a magnetic trap with oscillation
frequencies
$(\omega_x, \omega_y, \omega_z) = 2\pi \times (7.4, 0.8, 4.3)$~Hz, for
$z$ vertical.  For $^{87}$Rb, the photon recoil velocity $v_r = \hbar k/m$ is
$5.88$~mm/s and the fall time $t_1 = v_r/g$ is 0.6~ms.
We drop the atoms by turning off the trap current.  The current decay is
non-exponential, but reaches $1/e$ of its initial value after
160~$\mu$s with repeatability better than 1~$\mu$s.  
Due to the finite
turn-off time, the atoms take longer than time $t_1$ to reach a momentum
of $-\hbar k$, and we compensate for this by
delaying the first reflection pulse.  

The standing wave was produced by a home-built diode laser 
with a wavelength of 780.193 nm, 27 GHz blue 
of the $5S_{1/2} \leftrightarrow 5P_{3/2}$ laser cooling transition.
An acousto-optic modulator was used to control 
the optical intensity.  The light was then 
coupled into a single-mode fiber, which provided spatial filtering and 
pointing stability. The output 
from the fiber passed vertically through the vacuum cell and 
was retro-reflected from an external mirror to produce the standing wave.  
At this detuning, the expected loss due to spontaneous emission 
is $7\times 10^{-4}$ for the order-1 reflection 
and $2\times 10^{-3}$ for the order-2 reflection.
The beam was approximately Gaussian with a waist of 1~mm.

We investigated suspension by releasing the atoms and then applying
a sequence of reflection operations.  The time before the first pulse
and the time between pulses were varied to maximize the number of
atoms remaining; the final number of atoms and their momentum state was
monitored using time-of-flight absorption imaging with a resonant
probe traveling along the horizontal $y$-axis.  The results
for both order-1 and order-2 bouncing are shown in Fig.~2.  In both
cases, suspension times exceeding 100 ms were observed. 

\begin{figure}
\begin{center}
  \includegraphics[width=3in]{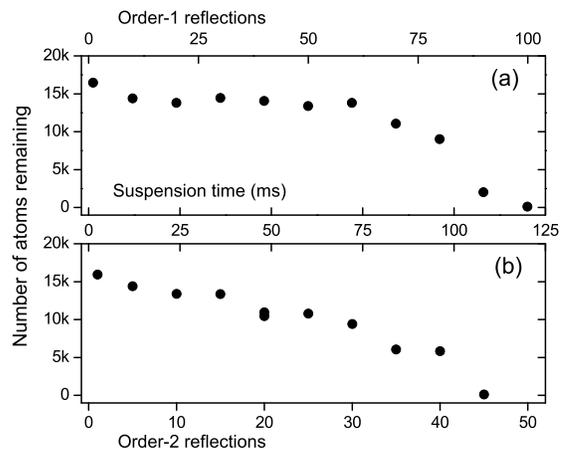}
  \caption{
  Suspension of atoms using (a) order-1 and (b) order-2 reflection operations.
  The top and bottom scales show the number of reflections used in the 
  respective experiments, while the center shows the total suspension time
  on a common scale.  Data points indicate the number of atoms
  remaining after the reflections.
}
\label{fig:bouncing_order1}
\end{center}
\end{figure}

We observe a non-exponential decay of the atom number,
indicating that atom loss is larger for the later operations.  The
form of this fall-off varied from day to day, but the time scale
was consistent.  The reason for the decay is not yet clear, but a few 
possibilities can be suggested.
For instance, if the reflection timing is incorrect, then the momentum error
$\hbar\delta$ will increase over time, leading to a reduction in 
operation fidelity.  Also, the condensate expands considerably in this time,
and if the standing-wave intensity is insufficiently uniform, 
spatially-dependent
errors in $\beta$ will develop.  We plan to investigate these issues
further, since according to our model, thousands of bounces should be 
possible.

The bouncing experiment already provides a measurement of gravity, 
since the optimum timing depends on the value of $g$ \cite{Impens06}.
From the bouncing
experiments shown, we obtain $t_1 = 603.0 \pm 0.5$~$\mu$s, where the
uncertainty is determined as the variation sufficient to 
reduce the atom number by roughly a factor of two.  
This gives $g = (\hbar k)/(m t_1) = 9.759
\pm 0.008$~m/s$^2$, noticeably different from the expected value of
9.81~m/s$^2$.  Since the atoms are in a state with non-zero magnetic 
moment, the discrepancy can be explained by a modest
ambient magnetic field gradient.  
Measurements outside the vacuum cell indicated
a vertical gradient of $B' = 7 \pm 2$~G/m.  
We also measured the gradient at the position
of the atoms by modifying the trap turn-off procedure to be non-adiabatic
for the atomic spins, so that multiple hyperfine states were populated.  As
the atoms bounced, the magnetic force caused the states to 
separate, as in a Stern-Gerlach experiment.  The separation could
be observed in the absorption images, and from it
we obtain a more accurate value of $ B' = 8.6 \pm 0.1$~G/m. This
gives a corrected value for $g$ of $9.814\pm0.008$~m/s$^2$.
We note that well-localized Stern-Gerlach measurements with long
interaction times are are
already an interesting application for the suspension technique.

A more precise determination of $g$ can be made
by implementing an atom interferometer using the suspended atoms.
One way to achieve this is illustrated in Fig.~1(c).  The atoms
are dropped as before, but at time $t_1$ when $p = -\hbar k$, the
beam-splitting operation $\ket{-\hbar k} \rightarrow 
\frac{1}{\sqrt{2}}\left(\ket{-\hbar k}-i\ket{+\hbar k}\right)$
is applied.
The two resulting wave packets can then be independently suspended using
alternating order-1 and order-2 reflections, as shown.  (Note that
the order-2 reflection has no effect on the momentum of atoms with
$p = 0$.)  After many reflections, the packets can be recombined
by another beam-splitting operation, with a result that depends on their
phase difference.  Since one packet is always above the other, it
is clear that the phase difference will be sensitive to $g$.

We calculate the phase difference between the packets using the 
quantum-mechanical solution for a falling plane wave state.  
A packet initially described by a wave function
$\psi(z,t=0) = \exp(iqz)$ will at subsequent times become
$\psi(z,t) = \exp[i(q-\gamma t)z] \exp[i\Theta(q,t)]$ for
$\gamma \equiv mg/\hbar$ and  
\be \label{eq-theta}
\Theta(q,t) = \frac{\hbar}{2m}\left(q^2 t - q\gamma t^2 
+\frac{1}{3}\gamma^2 t^3\right).
\ee
This can be verified by substitution into Schrodinger's equation.  
The phase difference developed during one cycle of the interferometer 
can therefore be expressed as
\be \label{eq-phi1}
\Phi = \Theta(k+\delta,2\tau) 
- \Theta(-k+\delta,\tau)-\Theta(3k-\gamma \tau+\delta, \tau) 
+ \phi_{r2} + \phi_{r1}
\ee
where $\hbar\delta$ is the momentum offset at the start of the cycle, 
$2\tau$ is the cycle
duration, and $\phi_{r_n}$ is the phase difference imparted by an order-$n$
reflection.  In addition, the momentum offset changes from
$\delta$ to $\delta - 2k + 2\gamma\tau$.
Evaluation of (\ref{eq-phi1}) yields
\be
\Phi = \frac{\hbar k}{m}\left(-4k\tau+2\gamma\tau^2\right) + 
\phi_{r1}+\phi_{r2},
\ee
Thus, after $N$ cycles, the wavefunction will be
\be
\psi = 
\ket{\hbar(k+\delta_N)} - ie^{iN\Phi}
\ket{\hbar(-k+\delta_N)}
\ee
for final momentum offset $\hbar \delta_N$.  The beam-splitting operation
is then applied with shifted phase $\phi_s$, 
resulting in a fraction of
atoms $f_+ = \sin^2[(N\Phi + \phi_s)/2]$ with momentum $+\hbar k$. 
We vary $\phi_s$ by shifting the frequency
of the standing-wave laser before the final beam-splitter, as in 
\cite{Garcia06}.
By plotting $f_+$ vs $\phi_s$, the phase $N\Phi$ is determined for
various numbers of cycles, 
as seen in Fig.~3.  We find $\Phi = 2\pi j-0.035 \pm 0.003$,
where $j$ is an integer.
The value of $g$ obtained from the bouncing experiments
fixes $j = -9$.

\begin{figure}
\begin{center}
  \includegraphics[width=3in]{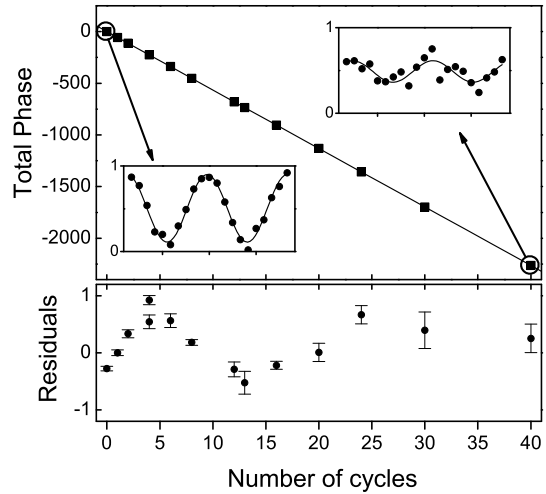}
  \caption{
  Results of interferometer experiment.  The upper graph shows the 
  total phase shift observed in the output, in radians.  The insets
  are sample plots of the fraction $f_+$ of atoms with $p = \hbar k$
  exiting the interferometer.  The cases of one cycle and forty cycles
  are shown.  
  The duration of one cycle is 2.4 ms.
  The residuals plotted in the lower graph are the difference between the 
  measured phase and the linear fit; the oscillations are
  attributed to a transient magnetic field produced when the trap is 
  turned off.
}
\end{center}
\end{figure}

The accuracy of the data is limited by deviations from the expected
linear dependence on $N$, as seen in the residuals in
Fig.~3(b).  We attribute the oscillating structure
to residual oscillations in the magnetic field after turning off
the trap.  The signal corresponds to a decaying gradient with initial amplitude
10~G/m, about 0.3\% of the original trap field gradient.  
Eliminating this field might
prove difficult, but its effects could be significantly reduced by
transferring the atoms to an $m=0$ hyperfine state before turning off the trap,
since their magnetic moment would then be nearly zero.

To analyze the measured phase, the reflection phases $\phi_{r1}$ and
$\phi_{r2}$ must be determined from the model.  
To do so accurately, the effect of gravity
during the pulse should be included.  This is accomplished by working in
the interaction picture with respect to the gravitational interaction
$mgz$.  The calculation proceeds as in Eq.~(\ref{eq-bloch}), but using
the interaction Hamiltonian
\be
H_I(t) = U_0^\dagger(t) V_L(z,t) U_0(t)
\ee
where $U_0(t) \ket{q} = \exp[i\Theta(q,t)] \ket{q-\gamma t}$.  We reference
$t$ to the beginning of the pulse sequence for the initial beam-splitter 
operation, to the center of the reflection sequences, and to the end
of the final recombination sequence.  The cycle time $\tau$ is defined 
accordingly.
We obtain $\phi_{r1} = (1\pm 1)\times 10^{-2}$ and
$\phi_{r2} = 0.56 \pm 0.16$.  These uncertainties
are the dominant source of error in the experiment.
They arise primarily from a sensitivity of the phase to the
intensity of the standing wave, which is difficult to control
precisely.  We obtain a value for $g$ of $9.745 \pm 0.027$~m/s$^2$, actually
less precise than that obtained from bouncing.  However, if the
reflection phase errors were 
eliminated, the fractional uncertainty would 
be reduced to $5\times 10^{-5}$.

The sensitivity to the standing wave intensity comes from a lack of 
symmetry between the two arms, since the lower arm undergoes 
an order-2 reflection 
while the upper arm does not, and gravity acts in the opposite sense for
the order-1 reflection.  This asymmetry can be removed if a
different beam-splitter operation is available, such as $\ket{0}
\rightarrow (\ket{0}+\ket{2\hbar k})/\sqrt{2}$.  This operation requires
a traveling standing wave, rather than the static standing wave used up
to now.  An interferometer
similar to the one described above could be implemented by following
the beam-splitter with
order-2 reflections at intervals of $t_2 = 2\hbar k/mg$.  Each packet
would undergo the same operations during a cycle so the reflection
phases would largely cancel.  In addition, the average packet separation
is twice as large, leading to a proportionally greater sensitivity to $g$.
We have implemented
an interferometer of this sort, but the beam-splitter was not sufficiently
consistent to achieve clear results.  We plan to implement a more reliable
beam-splitter using, for instance, the method of Ref.~\cite{Muller08}.

A traveling standing wave also permits more
complex interferometer schemes, such as that of Fig.~1(d).  This 
trajectory starts with the asymmetric beam-splitter described above.
A time $t_2$ later, an order-2 reflection is applied, making
$p = 0$ for the upper packet and $p = +2\hbar k$ for the lower packet.
This is immediately followed by a $\pi$-pulse on the transition
$\ket{0} \leftrightarrow \ket{+2\hbar k}$, leaving the upper trajectory
moving upward and the lower trajectory stationary.  Another time $t_2$ later, 
the same operations can be repeated, resulting in a controlled vertical
separation of the packets as shown.  When desired, the operations can be
reversed, bringing the packets back together and closing the interferometer
in a symmetric way.  Here 
the spacing between the packets increases with the total measurement time,
just as it does in an interferometer with free-falling atoms.  The fundamental
sensitivity of such a scheme would be comparable to that obtained with
free-falling atoms, but the minimal vertical space required would
be a substantial advantage.

The gravimetry techniques discussed here can be directly compared to
measurements using Bloch oscillations in an optical lattice, in which
the same type of ``bouncing'' transitions occur but in a lattice that
is continuously present \cite{Raizen97}.  
As a result, the momentum state of the
atoms oscillates at the Bloch frequency $\Omega = \pi mg/(\hbar k)$. 
By measuring $\Omega$, a value for $g$ can be obtained.  If the measurement
occurs over time $T$ and is shot-noise limited, then the accuracy 
$\delta\Omega$ will be 
$1/(T\sqrt{\mathcal{N}})$ for atom number $\mathcal{N}$.  
This gives an uncertainty $\delta g$ 
of $(v_r/\pi)(T\sqrt{\mathcal{N}})^{-1}$ for recoil velocity $v_r$.  
To compare, in our scheme the interferometer
phase can be measured to an accuracy of 
$1/\sqrt{\mathcal{N}}$, for an uncertainty
$\delta g = (\partial \phi/\partial g)/\sqrt{\mathcal{N}}$.  For the
interferometer of Fig.~1(c), this is 
$(g/v_r k)(T\sqrt{\mathcal{N}})^{-1}$, better than
the Bloch method by a factor of nine for the case of $^{87}$Rb.
A more complex interferometer scheme would give a
greater advantage, but the Bloch method could also be 
improved by the use of higher-order resonances \cite{Ivanov08}.

The authors of \cite{Impens06} point out that the simple bouncing experiment 
can be a good gravimeter if long-duration reflection operations are
used so as to 
maximize the velocity sensitivity.  The measurement precision 
depends on the details of the implementation, but scales favorably as
$T^{-3/2}$.

In conclusion, we have demonstrated the ability to suspend otherwise
unconfined atoms in gravity for over 100 ms, using repeated
reflections from a standing wave laser beam.  We note that free
atoms would fall a distance of 5~cm in this time.  To achieve this,
we developed high-precision reflection operations, for which our model
predicts that a factor of ten greater suspension time should be possible.
We further demonstrated a gravitationally sensitive atom interferometer
using the suspended atoms.  With some improvements, this technique might
be able to achieve gravimetric precision similar to that of free-falling
atom interferometers, but with a much reduced space requirement. 

We are grateful for helpful conversations with B.~Deissler, V. Ivanov,
and G.~Tino. 
We thank J.~Tiamsuphat and K.~Saenboonruang for contributions to the work.
This research was sponsored by the 
Defense Advanced Research Projects Agency (award 
No.~51925-PH-DRP) and by the National Science Foundation (award 
No.~PHY-0244871).

\bibliographystyle{apsrev}

\end{document}